\documentclass[journal]{IEEEtran}
\ifCLASSOPTIONcompsoc
  \usepackage[nocompress]{cite}
\else
  \usepackage{balance}
    \usepackage{cite}
\fi
\usepackage{graphicx}
\usepackage{amsthm}
\usepackage{amsmath}
\usepackage{amssymb}
\usepackage[ruled,vlined,linesnumbered]{algorithm2e}
\usepackage{smartdiagram}

\begin{document}
\title{Group Handover for Drone Base Stations}

\author {Yucel~Aydin,
        Gunes~Karabulut~Kurt,~\IEEEmembership{Senior Member, IEEE},\\
	~{Enver~Ozdemir,~\IEEEmembership{Member, IEEE},} and {Halim~Yanikomeroglu,~\IEEEmembership{Fellow, IEEE}  }  

\thanks{Y. Aydin is with the Institute of Informatics, Istanbul Technical University, Istanbul, 34485 {\color{black}Turkey} (email: aydinyuc@itu.edu.tr).}

\thanks{G. Karabulut Kurt was with the Department of Electronics and Communication Engineering, Istanbul Technical University, Istanbul, 34485, Turkey, when this work was performed. She is now with the Department of Electrical Engineering Polytechnique Montréal, Montréal, QC, H3C3A7, Canada (e-mail: gunes.kurt@polymtl.ca).}

\thanks{E. Ozdemir is with the Institute of Informatics, Istanbul Technical University, Istanbul, 34485 {\color{black}Turkey} (e-mail: ozdemiren@itu.edu.tr).}

\thanks{H. Yanikomeroglu is with the Department of Systems and Computer Engineering, Carleton University, Ottawa, K1S 5B6 {\color{black}Canada} (e-mail: halim@sce.carleton.ca).}

\thanks{Copyright (c) 2021 IEEE. Personal use of this material is permitted. However, permission to use this material for any other purposes must be obtained from the IEEE by sending a request to pubs-permissions@ieee.org.}

} 

\maketitle
\begin{abstract}
 The widespread use of new technologies such as the Internet of things (IoT) and machine type communication (MTC) forces an increase on the number of user equipments {\color{black}(UEs)} and MTC devices that are connecting to mobile networks. Inherently, as the number of UEs inside a base station's (BS) coverage area surges, the quality of service (QoS) tends to decline. The use of drone-mounted BS (UxNB) is a solution in places where UEs are densely populated, such as stadiums. UxNB emerges as a promising technology that can be used for capacity injection purposes in the future due to its fast deployment.
However, this emerging technology introduces a new security issue. Mutual authentication, creating a  communication channel between terrestrial BS and UxNB, and fast handover operations {\color{black}may cause} security issues in the use of UxNB for capacity injection. {\color{black}This} new protocol {\color{black}also} suggests performing UE handover from terrestrial to UxNB as a group. To the best of the authors' knowledge, there is no authentication solution between BSs according to LTE and 5G standards.  The proposed {\color{black}scheme} provides a solution for the authentication of UxNB by {\color{black}the} terrestrial BS. {\color{black}Additionally}, a credential sharing phase for each UE in handover is not required in the proposed method. The absence of a credential sharing step saves resources by reducing the number of communications between BSs. Moreover, many UE handover operations are completed in concise time {\color{black}within} the proposed group handover method.
\end{abstract}

{\color{black}
}
\IEEEpeerreviewmaketitle

\section{Introduction}
A substantial surge in the number of user equipments (UE) utilizing mobile services is expected {\color{black}in} the near future. As the number of UEs connected to a terrestrial base station (BS) increases, the quality of service (QoS) per user tends to reduce. It is highly probable that the BS will even be out of service, and therefore, UEs will not be able to {\color{black}access to} their mobile services. The current solution for such situations is applying to capacity injection, such as a mobile BS \cite{MobileBS}. The service provider deploys {\color{black}mobile BSs} in {\color{black}a} crowded area, which eventually increases the mobile network's average QoS.

A radio access node on-board {\color{black}of} unmanned aerial vehicle (UxNB) is a radio access node providing service to the UEs deployed on an unmanned aerial vehicle (UAV) according to 3GPP TS 22-125 \cite{3GPP22125}.  {\color{black} The UxNB can connect to the core network as} terrestrial base station, which is next generation NodeB (gNB) in 5G new radio (NR). The research community already has an interest in UxNB in order to enhance the mobile network coverage. {\color{black}The UxNB} can be exploited in several scenarios, such as emergencies, and high-density areas. UxNB can be deployed {\color{black}to} an area without terrain constraints\cite{3GPP22829}.

{\color{black}
Providing uninterrupted service to many different types of UEs is one of the main focus areas of 6G research activities \cite{DroneCell}. The traffic requirements expected from mobile networks may vary depending on the usage scenario. Mobile networks will need to be reinforced towards scenarios such as unforeseen natural disasters, traffic congestion, high-density concerts, or football games. Public safety communication systems are indispensable for rescue teams {\color{black}in case of disasters}. However, this infrastructure is also affected {\color{black}by} a disaster \cite{publicsafety}. In order to ensure the continuity of communication, current mobile network infrastructure should be rearranged in accordance {\color{black}of} such situations. Thanks to their assets and deployment advantages, UxNB is the main candidate to close these gaps in the current mobile networks. UxNBs can be exploited in high demand situations or public safety and disaster management operations. The main advantage of UxNB is the deployment capability to any area without {\color{black}an operating pilot}. For high-density areas, {\color{black}a} better QoS can be provided by UxNBs via capacity injection \cite{dronecell2}.
}

Security in mobile networks is often focused on the robustness of the UE authentication and encryption of over-the-air traffic. Only the air part of the mobile traffic is encrypted by UE and sent to the BS, and the traffic is decrypted in the BS and sent to the core network. Traffic within the core network is transmitted in plaintext. Traffic between base stations is therefore sent in plaintext. For example, during a handover operation, {\color{black}the} serving-BS (\textit{s-BS}) sends handover credentials to the target-BS (\textit{t-BS}) in plaintext. {\color{black}There is even no authentication process before this transmission} \cite{BSComm}. With fake BS attacks, attackers can capture personal data and track the location of UEs. The biggest reason for these attacks to be carried out is the lack of authentication within the core network. 

Fake BS attacks are carried out for International Mobile Subscriber Identity (IMSI) catching. A fake BS sends strong identity request signals to the UEs around it. UEs respond to this request by sending their IMSI information. Attackers who have captured IMSI information can authorize themselves through the nearest BS. The lack of authentication between BSs {\color{black}is the reason for such attacks. Thus, the attackers} can track all data and location of the relevant users. {\color{black} According to $2016$ figures, a professional fake BS production costs between $68,000-134,000$ USD, and a homemade {\color{black}forged} BS can be made for $1500$ USD \cite{IMSI}.} With the widespread use of UxNB technology in the future, this cost may decrease even more, and the UE data can be stolen in a short time with its fast deployment feature. There must inevitably be an authentication scheme between {\color{black}the} next-generation UxNBs and terrestrial BSs. Also, in order to encrypt the communication between {\color{black}two BSs}, both {\color{black}BSs} must have the same encryption key as a result of this authentication.

\begin{figure}[h!]
\centering
\includegraphics[width=\linewidth]{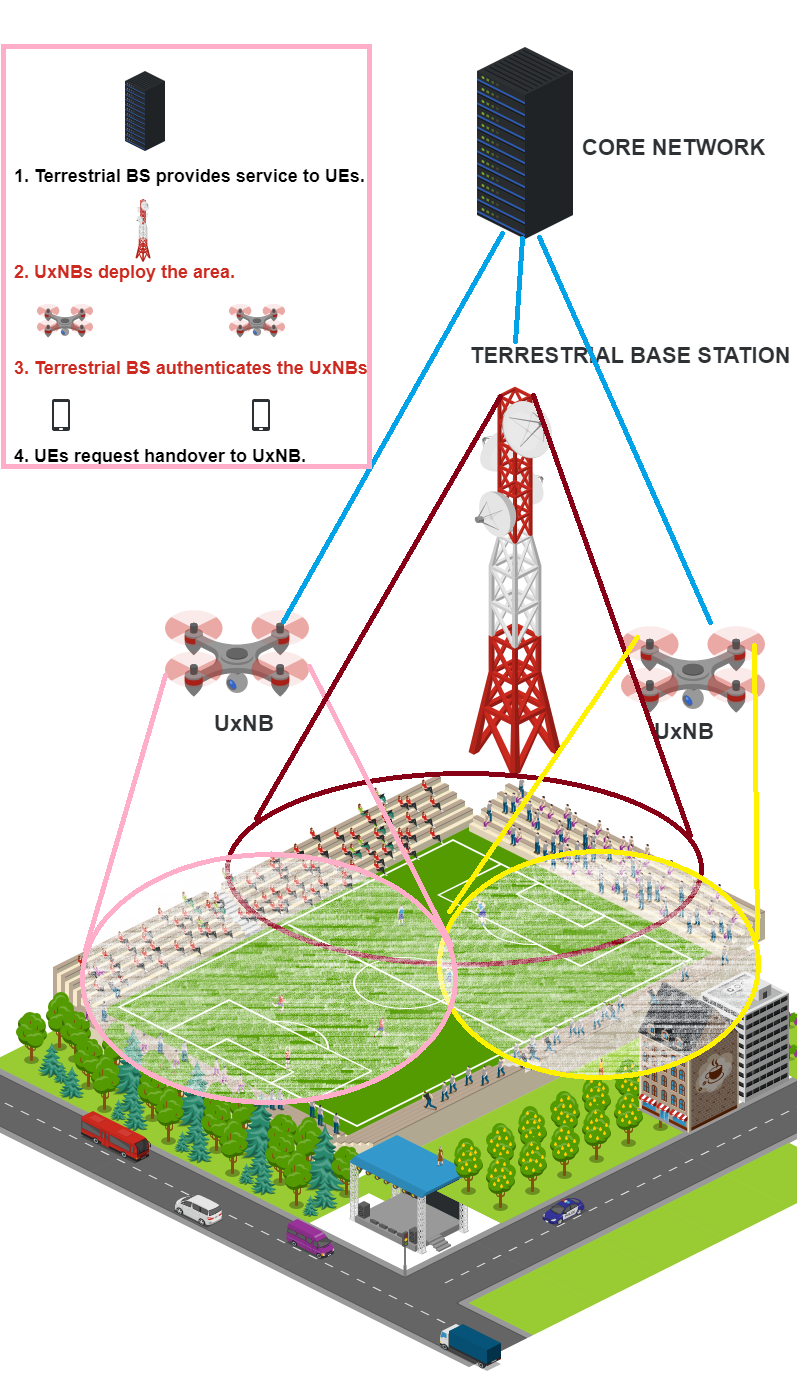}
 \caption{An examplary use case for capacity injection and group handover.\newline  The terrestrial BS cannot provide the required QoS and the UxNBs are sent to the game for capacity injection. The UxNBs should be authenticated by terrestrial BS and several handovers should be performed between terrestrial BS and UxNBs.}
\label{fig:usecase}
\end{figure}

As a use case, consider a football game where there is {\color{black}a steady} increase in the number of users in a particular area (stadium) in a specific period ($90$ min.). During the game, only one terrestrial BS may provide service to all UEs. It will be more beneficial to use UxNB to increase the BS's capacity, as shown in Figure \ref{fig:usecase}. While capacity injection is the first issue in such dense deployment scenarios, the handover among the overlay cells after capacity injection is the second issue. The use of several BSs within a particular area will result in overlay cells. Terrestrial BS providing service to all UEs in the stadium will delegate some UEs to UxNB to reduce the {\color{black}traffic load}. In the meantime, there will be many handover operations. In the currently used LTE \cite{3GPP36300}, and 5G standards \cite{3GPP501}, these handover operations should be done {\color{black}sequentially, {\color{black}i.e.} one by one. Yet, due} to the limitations of the UxNBs such as weight and battery life, their flying time will be a maximum of one hour \cite{3GPP22829}. Considering that a football game is {\color{black}at least} 90 minutes, {\color{black} a new UxNB will take over from ex-UxNB at least once}. This will cause an increase in the number of handover operations to be {\color{black}performed}. In this study, it is proposed that UEs should be transferred from terrestrial to UxNB not individually but as a group in order to make this handover process {\color{black}more efficiently}.

Group handovers usually take place in metropolitan cities, and {\color{black}the handovers occur} with the need for multiple UEs to be connected to another BS at the same time. A group of people inside a bus can be an excellent example of group handover. These UEs will receive a stronger signal from another BS at the same time as the signal strength from the \textit{s-BS} will decrease. Since it is the UE that chooses the BS to {\color{black}get service from}, many UEs will choose the new BS at the same time. This will cause bandwidth reduction in the new BS. At this point, a group handover problem will occur \cite{GroupHandover}. Handover of all UEs one by one, as in the standards, will cause service disruption and energy loss. The same problem is valid for the stadium example. When the UxNBs become active, many UEs will receive stronger signals from UxNBs and will request handover. An influential group handover solution is required in this scenario, as well.

In light of these challenges, our main contributions are listed as;
\begin{itemize}

\item {\color{black}A fast and energy-efficient handover scheme is proposed} for the high-density areas where UxNB can be exploited efficiently to provide service to the UEs. The current handover solution in both LTE \cite{3GPP36300} and 5G NR \cite{3GPP501} requires to share UE data from the \textit{s-BS} to the \textit{t-BS}. However, there is no data-sharing between BSs in our proposed method, {\color{black}which saves time and energy}. Besides, confirming UEs by the \textit{t-BS} as a group decreases the time for handover.

\item Fake BS attack is a security issue for current mobile networks. Using UxNB also creates the same problem. Any intruder can impersonate a UxNB and try to control UEs. Our proposal offers an authentication solution between BSs. UxNB can obtain public and private key pair from the core network before becoming active. When  UxNB comes over {\color{black}a} high-density area, the \textit{s-BS} can authenticate UxNB easily by using the public key of the \textit{t-BS} as explained in the group handover framework section.

\item As the \textit{s-BS} shares the group secret information with the \textit{t-BS}. By using this function, the \textit{t-BS} can authenticate UEs easily in the group handover phase. The authentication can be accomplished as a group. Thanks to having a private function, there is no phase for the control packet transmissions of data between BSs. While the \textit{s-BS} sends data for each UE to the \textit{t-BS} in 3GPP Release 16\cite{3GPP501}, no data-sharing between BSs is required in our proposed method. Therefore, in our proposed method, the number of control packet transmissions between BSs is zero.

\item In all authentication solutions for mobile networks, UE must have a private key. In the proposed method, {\color{black}it is} recommend that UE turns the private key into a public key with {\color{black}a powering operation in the elliptic curve group} for handover operation. Handover operation is carried out when the \textit{t-BS} verify the public key. These private keys must be distributed {\color{black}to UEs} before authentication. In the proposed method, it is possible to use a subscription permanent identifier (SUPI) belonging to each UE as a private key. This solution eliminates the need for private key distribution before authentication.
\end{itemize}

This paper is organized as follows. The next section provides an overview of the handover process in the 3GPP standards for both LTE and 5G NR and an overview of  existing handover methods. In Section III, the preliminaries of our proposal are explained in detail. System and thread models are given in Section IV. Our proposed approach for capacity injection and group handover is presented in Section V. The security and performance evaluation is provided in Section VI and Section VII, respectively. The study is completed by a conclusion in Section VIII. We present a list of abbreviations which are  used throughout the paper in Table \ref{table:abbreviation}.

\section{Literature Overview and Related {\color{black}Works}}

{\color{black} The section begins with the explanation of} the related works on mobile network handover with detailed information about handover both in LTE and in 5G NR. Although 5G NR will be the near-future standard of mobile networks, LTE has still been used by most countries. Ultra-densification is one of the new approaches for 5G NR. The coverage area of BSs is shrunk, and the number of users served by each BS is reduced. Therefore, the frequency of handover increases due to the densification's impact in 5G {\color{black}networks} \cite{LTEtoNR}. 

\subsection{UAVs in 3GPP Standards}
{\color{black}
According to 3GPP TS 22-261 \cite{3GPP22261}, it is predicted that UAVs are going to be used in several applications by governments and commercial sectors. Latency and reliability will be one of the first {\color{black}concerns for the} next generation 6G networks. UAVs will need more certain {\color{black}location} information and security {\color{black}against} theft and fraud.

An unmanned aerial system consists of UAVs and UAV controller \cite{3GPP23754}. The data traffic between these two components must be well-protected. Next generation mobile network providing service to the UAVs must be resistant to spoofing and non-repudation attacks.

Identification, tracking, and authorization of UAV and controllers are controlled by a central system, which is the Unmanned Aerial System Traffic Management (UTM) \cite{3GPP23754}. UTM stores all identity and meta information of UAVs and UAV controllers. The authentication and authorization of UAVs in the area have taken place by the information sharing procedures between UTM and mobile core network, especially AMF. It is clear that including of UAVs to the mobile network {\color{black}introduces a higher} computational burden for the AMF. 
}

\begin{table}[h!]
\caption{Abbreviations}
\label{table:abbreviation}
\centering
 \begin{tabular}{l l}
\hline
\textbf{Abbreviation} & \textbf{Description} \\
 \hline
3GPP & 3rd Generation Partnership Project \\
BS & Base Station \\
RAN & Radio Area Network \\
UE & User Equipment \\
NR & New Radio\\
QoS & Quality of Service \\
UAV & Unmanned Aerial Vehicle \\
s-BS & Serving Base Station \\
t-BS & Target Base Station \\
LTE & Long Term Evolution \\
MME & Mobility Management Entity \\
MR & Measurement Report \\
AMF & Access and Management Function \\
SGW & Serving Gateway \\
UPF & User Plane Function \\
SUPI & Subscription Permanent Identifier \\
ECDLP & Elliptic Curve Discrete Logarithm Problem \\
SUCI & Subscription Concealed İdentifier \\
UDM & User Data Management \\
AUSF & Authentication Server Function \\
SEAF & Security Anchor Function \\
MAC & Message Authentication Code \\
ENC & Encryption \\
gNB & Next Generation NodeB \\
UTM& Unmanned Aerial System Traffic Management \\
\hline
\end{tabular}
\end{table}

The use of UxNB to increase the coverage area is specified in 3GPP standards. A UxNB can connect to 5G core network as a terrestrial BS via wireless backhaul link\cite{3GPP22829}. A UxNB can be used in various scenarios such as emergencies, temporary coverage for UEs, hots-spot events, due to their fast deployment and broad coverage capabilities \cite{3GPP22829}. UxNBs should be authenticated by the core network before operating as a BS. One of the requirements for using a BS on UAV is to keep the energy {\color{black}usage} at the lowest level because UAV has limited {\color{black}power}. 

The use of UAVs alone is limited due to their airborne time and energy constraints. For example, using a single drone in delivery services results in waiting for that vehicle to come back to the base. For this reason, UAVs should be used as a swarm. The essential requirement for a swarm of UAVs is group management \cite{3GPP22829}. Group management requires group authentication and secure communication inside a group, as {\color{black}given in the following sections.}

\subsection{Handover Management in LTE}
There are two types of handover scenario in LTE {\color{black}based on} the existing of the mobility management entity (MME) change \cite{LTEtoNR}. Inter-BS with intra-MME is {\color{black}described} in this section step by step. The  UE disconnects from serving BS (\textit{s-BS}) and connects to the target (\textit{t-BS}) without changing MME. The reason we choose this scenario is that the number of communications is less than in other scenarios. 

\begin{figure}[h!]
\centering
\includegraphics[width=\linewidth]{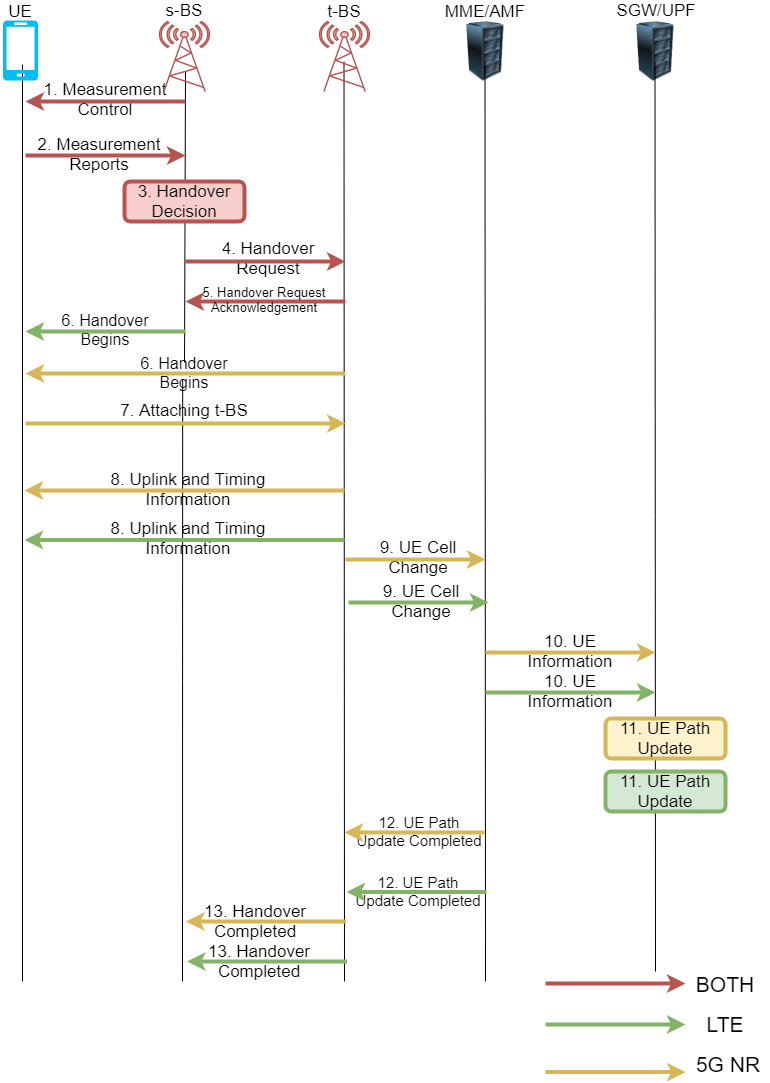}
 \caption{Handover in LTE and 5G NR. In LTE, handover begins with the \textit{t-BS} command to UE, whereas handover begins with the \textit{s-BS} command in 5G.}
\label{fig:ltehandoff}
\end{figure}

The handover steps are shown in Figure \ref{fig:ltehandoff}{\color{black}; and also listed below.}

\begin{enumerate}
	\item The UE measurement procedure is configured by the \textit{s-BS}.
	\item The UE sends  a measurement report (MR) to the \textit{s-BS}.
	\item According to the report, the \textit{s-BS} makes a handover decision.
	\item The \textit{s-BS} sends a handover request to the the \textit{t-BS}.
	\item The \textit{t-BS} sends an acknowledgment to the \textit{s-BS} according to its resources.
	\item The \textit{t-BS} informs the UE for handover with necessary information.
	\item The UE attaches to the target cell.
	\item The \textit{t-BS} sends uplink allocation and timing information to the UE.
	\item The \textit{t-BS} informs the MME for UE cell change.
	\item MME informs the serving gateway (SGW) for UE.
	\item SGW updates the path for UE.
	\item MME informs the \textit{t-BS} for path update.
	\item The \textit{t-BS} informs the \textit{s-BS} for the completion of the handover.
\end{enumerate}

\subsection{Handover Management in 5G NR}
The handover procedure for 5G NR is almost the same with LTE with little changes. Access and mobility management function (AMF) executes the duties of MME, while the user plane function (UPF) is the same as SGW.

The handover steps are listed as:

\begin{enumerate}
	\item The UE measurement procedure is configured by the \textit{s-BS}.
	\item The UE sends  MR to the \textit{s-BS}.
	\item According to the report, the \textit{s-BS} makes a handover decision.
	\item The \textit{s-BS} sends a handover request to the \textit{t-BS}.
	\item The \textit{t-BS} sends an acknowledgment to the \textit{s-BS} according to its resources.
	\item The \textit{s-BS} sends a handover command to the UE.
	\item The UE attaches to the target cell.
	\item The \textit{t-BS} sends uplink allocation and timing information to the UE.
	\item The \textit{t-BS} informs the AMF for UE cell change.
	\item AMF informs UPF for UE.
	\item UPF updates the path for UE.
	\item AMF informs the \textit{t-BS} for path update.
	\item The \textit{t-BS} informs the \textit{s-BS} for the completion of the handover.
\end{enumerate}

\subsection{Key Hierarchy in 5G NR}
\begin{figure}[h!]
\centering
\includegraphics[width=\linewidth]{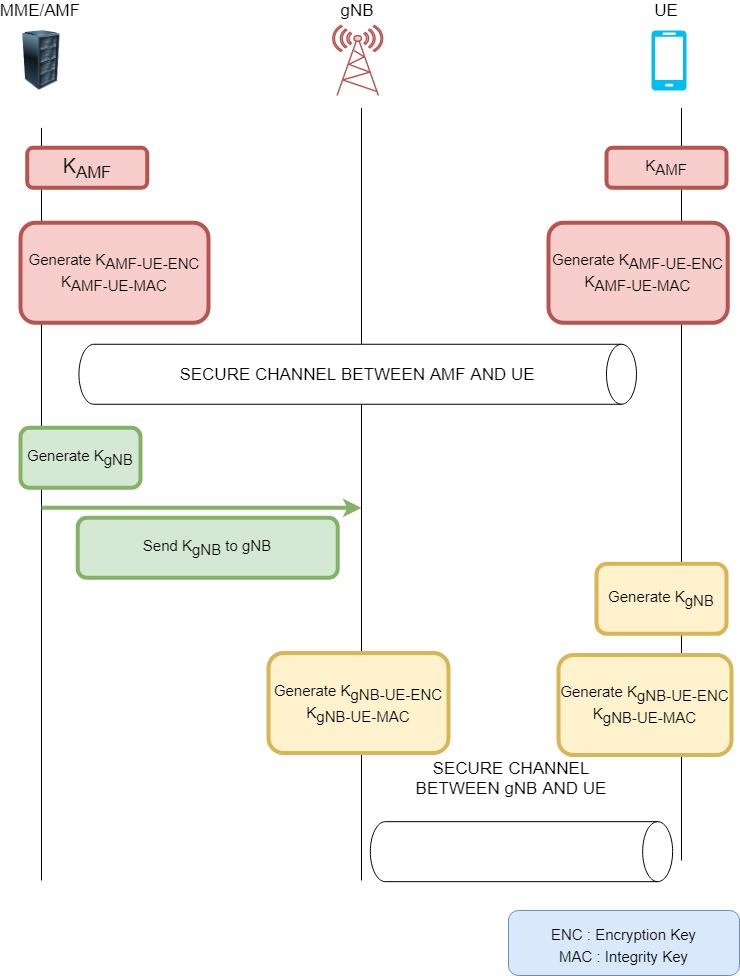}
 \caption{Key Generation in 5G NR. $K_{gNB}$ is derived from $K_{AMF}$. Both AMF and UE {\color{black}have} the $K_{AMF}$. AMF sends $K_{gNB}$ to BS. UE can create $K_{gNB}$ for a secure communication with BS. The encryption and integrity keys are dervied from the main keys $K_{AMF}$ and $K_{gNB}$.}
\label{fig:keyhierarchy}
\end{figure}

It is crucial to understand the key generation process in 5G NR in order to explain the key exchange between the base stations in the handover phase. The key generation steps are {\color{black}depicted} in Figure \ref{fig:keyhierarchy}. The main key ($K_{AMF}$) is known by AMF and UE \cite{vulnerabilities}. The key for BS ($K_{gNB}$) and the integrity and encryption keys ($K_{AMF-UE-INT}$, $K_{AMF-UE-ENC}$) for the secure communication between AMF and UE are derived from $K_{AMF}$. AMF sends $K_{gNB}$ to BS, and UE can compute the  same keys since UE has the primary key. Once BS has the $K_{gNB}$, the integrity and encryption keys {\color{black}($K_{gNB-UE-INT}$ and $K_{gN-UE-ENC}$, respectively)} for the secure communication between BSs and UE can be computed by BS.

\subsection{Handover Key Exchange in 5G NR}
Once the handover decision is taken by the \textit{s-BS}, the BS key ($K_{gNB}$) is shared with the \textit{t-BS} \cite{vulnerabilities}. The \textit{t-BS} computes the next BS key value $K_{gNB*}$ by using $K_{gNB}$ and other security parameters, which are the same for all BS connected to the same AMF. The new integrity and encryption keys for secure communication between UE and the \textit{t-BS} are derived from $K_{gNB*}$. The UE also can compute $K_{gNB*}$, since the UE has $K_{gNB}$ and security parameters. Then, the UE can compute the new encryption keys and message authentication codes (MAC) for further communications.

{\color{black}
\subsection{Security Exploits in 5G NR}
Most of the vulnerabilities in LTE-A, such as IMSI-catching are {\color{black}addressed} in 5G NR. However, there are still on-going security issues in 5G NR \cite{exploits}. SUPI is the encrypted version of IMSI in 5G. IMSI, which is the UE's private key, is sent to BSs as cleartext in the LTE-A standards. The UE's private key in 5G NR (SUPI) is encrypted by using the public key of the BS. Unfortunately, with the fake BS attack, the attackers can still obtain the UE's private key. 

According to 3GPP Release 16\cite{3GPP33846}, attackers can perform SUPI guessing attacks \cite{SUPIGUESS}. Attackers can generate random SUPI and encrypt it by the BS public key. If the attackers obtain a valid response from BS, the SUPI can be assumed as a real SUPI. 

Authentication complexity is another security issue for 5G handover process. The assumption of extra mobility in 5G can increase authentication complexity. {\color{black} {\color{black}BSs} have to share UE data between each other and the core network must be informed as well. {\color{black} The surge in mobility rate results in increasing number of communications.} Therefore, mobility is one of the most important aspects that affect authentication complexity \cite{vulnerabilities}.  }

\subsection{Handover Studies for UxNBs}
There are few studies on the security aspect of UxNB since the usage of UxNB is a new emerging topic. In \cite{Wifibackhaul}, the authors investigate the scenario in which UEs move from one UxNB to another one. The difficulty of using the $X_2$ logical interface as in LTE  is explained in the study. 

The studies \cite{handoverdecision1,handoverdecision2} are related to the selection the best time for the handover. The aim of the papers is to decrease the handover latency. The security aspect of the handover process is not taken into account. 

There exist some researches which investigate the security link between a UAV and control station and the handover key management in LTE-based aerial vehicle control network \cite{UAVControl1, UAVControl2}. {\color{black} In the studies, authentication between UAV and BS is accomplished by key pairs instead of IMSI. With key pairs, which UAV and BS already know before communication, the main authentication key is created and authentication and handover are performed with the new authentication key.} According to {\color{black}the presented} simulation results, the handover of UAVs is mostly performed between BSs. 
}

One of the gaps in the previous studies is the lack of the authentication solution between UxNB and terrestrial BS. An intruder can impersonate a UxNB and be involved to the system without authentication. Also, the scalability problem for terrestrial BSs is not addressed in the studies. In high-density areas, BSs can drop the requests from UEs. Besides, the scalability is also an issue for handover in high-denstiy areas. These points are not taken into consideration in the previous studies. Our proposed scheme is based on capacity injection for the densely populated areas and group handover and addresses the authentication issue between BSs. Besides, the scheme is a promising solution for handover between terrestrial BS and UxNB by using less energy and time for scalability. {\color{black}This study {\color{black}is based on the} previous work \cite{arXiv1}. In \cite{arXiv1}, a group authentication scheme was proposed to authenticate users in a group at the same time. The group handover method appears as an application area of the proposed group authentication scheme. While all users in a group are verified for the first time in the group authentication scheme, in the group handover solution, there {\color{black}exists one} BS, which already authenticated the UEs in its coverage area by the group authenticatin scheme and a sub-group of these UEs will be handed over to the UxNBs coming to the area for capacity injection.}

\begin{table}[h!]
\centering
{\color{black}
\caption{Notation}
\label{table:notations}
 \begin{tabular}{l l}
\hline
\textbf{Notation} & \textbf{Description} \\
 \hline
$G$ & A Cyclic Group \\
$P$ & A Generator on $G$ \\
$E$ & Encryption {\color{black}Function} \\
$D$ & Decryption {\color{black}Function} \\
$t$ & Threshold Value\\
$s$ & Private key \\
$H(\cdot)$ & Hash Function \\
$f(x)$ & A Random Function  \\
$x_i$ & Public Key for Each User $U_i$ \\
$f(x_i),f(x_i)\cdot P$ & Private Key for Each User $U_i$ \\
$ID_i$ & Identification Number for Each User $U_i$ \\
$x$ & The Number of Emerging UxNB \\
$y$ & The Number of UEs in A Group \\
$A$ & Addition Operation \\
$ECA$ & Elliptic Curve Addition Operation \\
$ECP$ & Elliptic Curve {\color{black}Power} Operation \\
\hline
\end{tabular}
}
\end{table}

\section{PRELIMINARIES}

This section aims to provide a summary of building blocks employed in the proposed algorithm. {\color{black} The notation used in this paper is given in Table \ref{table:notations}. The} proposed scheme is built on the assumption that the terrestrial BS and UEs create a group and perform a group authentication as in the \cite{arXiv1}. In this section, we give the details of group authentication {\color{black}based on} the elliptic curve discrete logarithm problem (ECDLP), which is widely used in the proposal. 

The group authentication is based on the secret sharing scheme and ECDLP. The group authentication scheme consists of two steps. In the initialization phase, group manager (GM), which has better sources than other group members, selects the initialization parameters. These parameters are elliptic curve parameters and secret and public keys for each user. The GM determines these parameters and shares them with the relevant group members. The confirmation phase is to check whether the group members are legitimate. These phases stated below;

\begin{center}
  {\bf The Initialization Phase}
\end{center}

	GM selects a cyclic group $G$ and a generator $P$ for $G$. GM selects $E=Encryption(\cdot)$ and $D= Decryption(\cdot)$ algorithms and a hashing function $H(\cdot)$. A polynomial with degree $t-1$ is chosen by GM and the constant term is determined as {\color{black}the} group key $s$. GM selects one public key $x_i$ and one private key $f(x_i)$ for each user in the group $U$ where each user is denoted by $U_i$ for $i=1,\ldots,n$. GM computes $Q=s\times P$. GM makes $P, Q, E, D, H(s), H(\cdot), x_i$ public and shares $f(x_i)$ with only user $U_i$ for $i=1,\ldots,n$.

\begin{center}
{\bf The Confirmation Phase}
\end{center}

	The steps of the confirmation phase are performed as in the Algorithm 1. Each user computes $f(x_i)\times P$ and sends  $f(x_i)\times P\ ||ID_i$ to the GM and other users ($ID_i$) is the identification number of the user and $\|$ symbol shows the concatenation of two values). If the GM verifies the authentication, the GM computes $f(x_i)\times P$ for each user and verifies whether the values are valid or not. If the GM is not included in the verification process, any user in the group computes 
\begin{equation}
C_i=\left(\prod^{m}_{r=1, r\neq i}\dfrac{-x_r}{x_i-x_r})\right)f(x_i)\times P
\end{equation}
for each user ($m$ denotes the number of the users in the group and $m$ must be equal to or larger than $t$). User verifies whether
\begin{equation}
\sum_{i=1}^{m}C_i  {\stackrel{?}{=}}  Q
\end{equation}
holds. If $(2)$ holds, {\color{black}the} authentication is completed. Otherwise, the process will be repeated from the initialization phase.

\begin{algorithm}
{Compute  $f(x_i)\times P$, and share  $f(x_i)\times P \| ID_i$}\newline\\
\If{GM verifies the authentication}
{
	GM computes $f(x_i)\times P$.\newline\\
	\If{All values are valid}
	{
		Print `Authentication is complete.'
	}
	\Else{
		Repeat.
	}
}
\Else{
	{User computes $c_i$=$f(x_i)\times P{\overset{m}{\underset{r=1, r\neq i}{{\displaystyle\prod}}}(-x_r/(x_i-x_r))}$.}\newline\\
	\If{${\overset{m}{\underset{i=1}{{\displaystyle\sum}}}c_i}$ = Q}
	{
		Print `Authentication is complete'.
	}
	\Else{
		Repeat.
	}
}
\BlankLine
\caption{Confirmation Phase}
\end{algorithm}
{\color{black}
Subscription concealed identifier (SUCI) and SUPI are the keys used in the initial authentication between UE and BS, according to the 3GPP Release 16\cite{3GPP501}. These globally unique identifiers can be {\color{black}used} both in group authentication between BS and UEs and in our handover proposal. As explained in the group authentication scheme, each UE must have a private key $f(x_i)$ to participate in the group authentication. In \cite{arXiv1}, these private keys are shared with UEs in the initialization phase. For the proposal, $x_i$, which are public, can be equal to the encrypted version of a globally unique SUPI or SUCI.

In the} proposal, ECDLP is significantly used. Given an elliptic curve over a finite field $F_p$ and two points $P,Q$ over {\color{black}the} elliptic curve, to find an integer $k$ such that $Q = kP$ is defined as a discrete logarithm problem. {\color{black}The ECDLP is exploited} in order to provide confidentiality of the private keys $f(x_i)$ by multiplying it with a point $P$ to obtain a new point $f(x_i)P$.

\section{System and Threats Models}

In this section, details of capacity injection and group handover scenario in a densely populated area where the proposed method can be used efficiently are given. In addition, the attacks that can be performed against the specified scenario are described.

\subsection{System Model}
A terrestrial BS provides service to a group of UEs in its coverage area in the system model. Due to the number of the UEs and the {\color{black}corresponding} high volume of traffic, the service provider sends UxNB to the close area of the terrestrial BS. Both terrestrial and UxNBs have secure connectivity with AMF. The UxNBs should be authenticated by the terrestrial BS. Afterward, the UEs in the range of UxNB should be handed-over from terrestrial to UxNB. The UxNB may confirm the UE by sending the handover request to the AMF for each UE. However, this can be time-consuming, or AMF can collapse due to the number of UEs. A fast and lightweight handover scheme is needed for the system.

According to group authentication solutions, a certain number of devices with the same or different capacities come together to form a group in order to perform a fast authentication. Within this group, members are preferred to authenticate quickly as a group rather than authenticate with each other one by one. In group authentication methods, the initial parameters must be determined by a GM. The GM generally has more resources and capacities than other members. In the determined system model, the GM is the terrestrial BS that provides service to the UEs. The terrestrial BS and UEs inside the stadium form a group. 

\subsection{Threat Model}
The communication between terrestrial and UxNB is vulnerable to man-in-the-middle and replay attacks. A malicious UxNB can interrupt the communication and authenticate itself by using the credentials taken from the legitimate UxNB. A secure authentication method is needed to detect legitimate and malicious UxNB. A malicious UE also can perform a man-in-the-middle attack to authenticate itself.

Attackers can eavesdrop on the traffic between UEs and BS if the communication is not encrypted. If any credential is in plaintext format, attackers even can impersonate UEs. {\color{black}In addition}, a fake BS attack can deceive the {\color{black}legitimate} UEs in order to capture the signals. Overall, the {\color{black}messages} must be transmitted as encrypted ciphertext, and source and target authentication must be performed.

\section{Capacity Injection and Group Handover Framework}
The secure capacity injection and group handover solution is {\color{black}described} in this section for the predetermined scenario. An UxNB should be authenticated by the closest terrestrial BS in order to assume the emerging UxNB is legitimate. After succeeding authentication, the handover of UEs, which are in the range of UxNB, must be fulfilled from terrestrial to UxNB. Before authentication of UxNB, we assume that terrestrial BS with UEs {\color{black}in certain} range formed a group and a group authentication was carried out as in the \cite{arXiv1}. Consequently, {\color{black}the} terrestrial BS has a function $f(x)$, which is private and only known by {\color{black}the} terrestrial BS and AMF. AMF must have a table which stores the identity of terrestrial BSs with their corresponding private function. {\color{black}In addition, after a} successful group authentication, each UE in the range of terrestrial BS has a private value $f(x_i)$ and public values $(x_i, f(x_i)P)$. The $i$ is the identity of UE, and $P$ is the generator in the elliptic group, which is used to keep $f(x_i)$ private by {\color{black}powering operation in the elliptic curve group.} AMF stores the private values of UEs in the database as well. To authenticate the new emerging UxNB, the work sequence at below should be followed.

\subsection{Authentication of Emerging UxNB for Capacity Injection}

	AMF assigns private key $f(x_i)$ and public key pairs ($x_i, f(x_i)P$) which did not designate any other UE to the UxNB as in Algorithm 2. If an adversary can obtain private keys $f(x_i)$ more than the threshold value, the secret function $f(x)$  can be recovered by an adversary and a legitimate UxNB can be impersonated.  Once UxNB is the range of terrestrial BS, UxNB transmit $x_i$ and $f(x_i)P$ pairs to terrestrial BS. Afterward, terrestrial BS verifies the pairs by {\color{black}using} the private function $f(x)$. Finally, if the UxNB is legitimate, $f(x)$ is shared with UxNB. Both terrestrial BS and UxNB have the private key $f(x_i)$ of UxNB. By a symmetric key encryption method, the {\color{black}function $f(x)$} can be encrypted and sent securely to the UxNB by terrestrial BS.

\begin{algorithm}
{Assign the private key $f(x_i)$ and public key $x_i, f(x_i)P$ to the UxNB. }\newline\\
{Transmit $x_i$ and $f(x_i)P$ pairs to terrestrial BS.}\newline\\
\If{$x_i$ and $f(x_i)P$ pairs are valid}
{
	Send $f(x)$.
}
\Else{
	{Not valid UxNB.}
}
\BlankLine
\caption{Authentication of Emerging UxNB for Capacity Injection
}
\end{algorithm}

After accomplishing of authentication of UxNB, BSs can communicate with each other confidently by a symmetric key encryption. After successful authentication of UxNB, group handover can be performed anytime needed. UEs send their public values ($x_i, f(x_i)P$) to UxNB and UxNB confirms UEs. The work sequence for group handover should be followed, as detailed below:

\subsection{Group Handover}

	Each UE sends its public value ($x_i, f(x_i)P$) to the UxNB as in Algorithm 3. UxNB performs addition operation for each $f(x_i)$ and $f(x_i)P$ separately. At the end of the additional computation, the total $f(x_i)$ value is multiplied by the generator $P$. If the result is equal to the total $f(x_i)P$ value, all UEs are valid. Otherwise, the UEs are verified one by one, as shown in Algorithm 3. After successful control, UxNB begins to provide service for UEs. All requests from UE to UxNB {\color{black}are} going to be encrypted by the private key $f(x_i)$ of UE, and also $x_i$ value should be appended to all data.

\begin{algorithm}
{Send ($x_i, f(x_i)P$) to UxNB.}\\
{$m$ is the number of UEs in the handover group.}\\
{$TotalX$ is equal to $0$.}\\
{$TotalPoint$ is the {\color{black}identity of the elliptic curve group.}}\\
\For{$m$ down to 1}
{		
	{$TotalX=TotalX+f(x_i)$.}\\
	{$TotalPoint=TotalPoint+f(x_i)P$.}\\
}
\If{$TotalPoint$ is equal to $TotalX.P$}
{
	{Provide service to the UEs.} \\
}
\Else{
	\For {$m$ down to 1}
	{
		\If{$f(x_m)P$ is valid}
		{
			{Provide service $UE_m$.}\\
		}
		\Else{
			{The UE is not valid.}\\
		}
	}
}
\BlankLine
\caption{Group Handover}
\end{algorithm}
{\color{black}
\subsection{Computational and Communication Complexity}
The proposed scheme consists of two stages. In the first stage, the UxNB authentication stage, UxNB sends the public key pair $(x_i,f(x_i)P)$ to the terrestrial BS {\color{black}in} the first transmission. {\color{black}In} the second transmission, the terrestrial BS sends acknowledgment to the UxNB for each UxNB. Therefore the communication complexity of {\color{black}the} first stage is proportional {\color{black}to} the number of emerging UxNB. {\color{black}The terrestrial BS} performs one {\color{black}powering operation in the elliptic curve group} $(f(x_i)P)$ for each UxNB in order to compare the value sent by UxNB. 

In the second stage, the group handover stage, UE sends its public key pair $(x_i,f(x_i)P)$ to the UxNB. Communication complexity is proportional {\color{black}to} the number of UEs. UxNB performs one addition operation $(TotalX+f(x_i))$ and one elliptic curve addition operation ($TotalPoint+f(x_i)P$) for each UE, and one {\color{black}powering operation in the elliptic curve group} to compare the end result as reported in Table \ref{table:comlexity}.

\begin{table}[h!]
\centering
{\color{black}
\caption{Computational and Communication Complexity}
\label{table:comlexity}
 \begin{tabular}{l l l }
\hline
\textbf{Stage} & \textbf{Computational} & \textbf{Communication} \\
\textbf{} & \textbf{Complexity} & \textbf{Complexity} \\
 \hline
UxNB Authentication & $x$ ECP & $2x$ \\
Group Handover & $y$ A, $y$ ECA , $1$ ECP & $1y$ \\
\hline
\end{tabular}
}
\end{table}
}

\section{Security Analysis}
{\color{black}The} security provided by our proposal for all possible attacks against the scenario {\color{black}which} we explained in the models section are explained one by one. The prevention for an attack is given as a theorem and the solution is proved in the proof for each attack scenario.
\\

\textbf{Theorem 1:} \textit{The malicious UxNB which captures public values ($x_i,f(x_i)P)$ of legitimate UxNB can not perform replay attack.}

\textit{Proof.} The secret function $f(x)$ is encrypted by the private key $f(x_i)$ of UxNB and sent it to the UxNB by terrestrial BS. The malicious UxNB may intercept the communication between legitimate UxNB and terrestrial BS. Also, it may pass the confirmation phase by sending public values of legitimate UxNB. However, when the terrestrial BS sends the $f(x)$ polynomial in the encrypted version, the malicious UxNB can not decrypt it due to not having the private key $f(x_i)$. {$ \hspace{0.6 cm} \qed$}  

\textbf{Theorem 2:} \textit{The malicious UE, which captures public values ($x_i,f(x_i)P_2)$ of UE, can not perform a replay attack.}

\textit{Proof.} The malicious UE may intercept the communication between legitimate UxNB and UE. Also, it may pass the confirmation phase by sending the public values of UE. However, when it requests service from legitimate UxNB, it can not send an encrypted message due to not having the private key $f(x_i)$. {$ \hspace{5.6 cm} \qed$} 

\textbf{Theorem 3:} \textit{The attacker who captures the value of $P$ and $f(x_i)P$  sent by the UxNB publicly cannot have knowledge of private key $f(x_i)$.}

\textit{Proof.} Given two points $P$ and $f(x_i)P$ on an elliptic curve group, it is hard to find the $f(x_i)$ value that provides the relationship  $f(x_i)P = f(x_i)\times P$. This open problem is called ECDLP. Therefore, it is hard to find $f(x_i)$ by having $P$ and $f(x_i)P$. {$ \hspace{6.9 cm} \qed$} 

\textbf{Theorem 4:} \textit{The attacker can not decrypt the communication between UEs and UxNB after handover.}

\textit{Proof.} The communication is encrypted by the private of UE, $f(x_i)$, and the public value $x_i$ is sent with the encrypted message. The UxNB can compute $f(x_i)$ with the knowledge of the secret function $f(x)$ and the UE public key $x_i$. Attackers can only eavesdrop on encrypted traffic. {$ \hspace{2.9 cm} \qed$} 
  
The attacks in the four theorems are very crucial problems for the security in the specified densely populated scenario. First, the intruder can act as a legitimate UE or UxNB by eavesdropping the communication channel and obtaining some security parameters. Keeping secret of private keys in the proposed system is very critical. If intruders can capture private keys equal to the specified threshold value, they {\color{black}can} recover the secret polynomial and decrypt the information in the entire system.

\section{Performance Analysis}

\begin{figure}[h!]
\centering
\includegraphics[width=\linewidth]{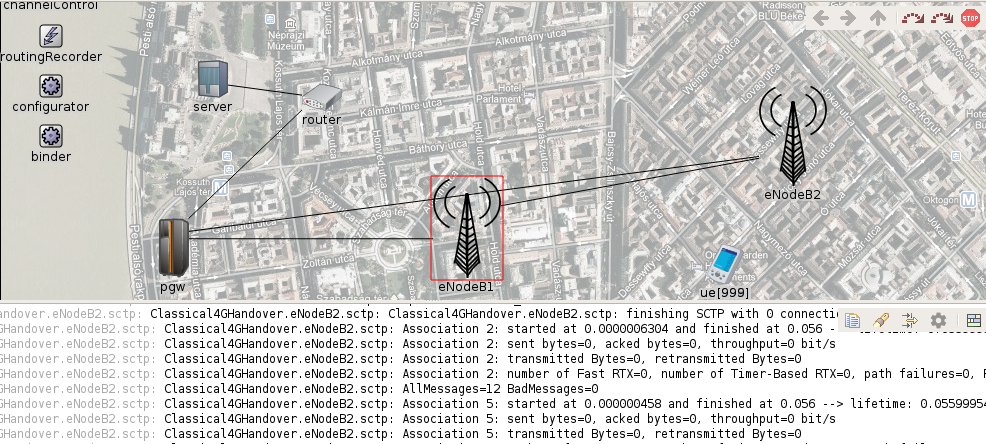}
 \caption{Simulation scenario. Several UEs detach from the \textit{s-BS} and attach to the \textit{t-BS}. Handover solution both in LTE and our proposed method is simulated by SimuLTE.}
\label{fig:omnetScenario}
\end{figure}

{\color{black}The} main objectives in the performance analysis are to show the importance of capacity injection for QoS and compare the handover time and the number of control packet transmissions in group handover. The SimuLTE\cite{simulte} library built on top of the Omnet++ {\color{black}package version 5.5.1} and INET framework {\color{black}are} used to simulate {\color{black}the} football stadium scenario, as seen in Figure \ref{fig:omnetScenario}. The most complex LTE scenarios can be simulated in SimuLTE in accordance with the 3GPP Release 16 \cite{3GPP36300}. The simulation framework exploits the layer base structured environment, and the handover process is accomplished mostly by the physical layer. Further, the $X_2$ link between BSs and protocols are well-designed and implemented by SimuLTE. 

{\color{black}
The difference between the proposed scheme and LTE handover solution is that there is no data sharing between BSs in our proposal. Therefore, the $X_2$ interface parameters should be taken into consideration carefully while performing simulation. Ethernet connection is used for the $X_2$ interface in SimuLTE simulation environment. Ethernet connection capacity is selected as {\color{black}$10$ Gbps} for our {\color{black}simulations}. With this connection capacity, data transfer between base stations is completed in {\color{black}$100$ ns.} Different configuration settings can be seen in Table \ref{table:datatransfer}. Other simulation parameters are selected as default parameters provided by SimuLTE.
\begin{table}[h!]
\centering
{\color{black}
\caption{Data Transfer Time between BSs}
\label{table:datatransfer}
 \begin{tabular}{l l}
\hline
\textbf{$X_2$ Ethernet Type} & \textbf{Data Transfer Time} \\
 \hline
$10$ Gbps & $100$ ns\\
$1$ Gbps & $522$ ns\\
$100$ Mbps & $14170$ ns\\
$10$ Mbps & $57250$ ns\\
\hline
\end{tabular}
}
\end{table}
}

{\color{black}
In the simulations, there are two BSs, core network and the UEs whose number can be changed to figure out handover time and the number of transmissions. UEs are placed at {\color{black}a} point between BSs where the handover operation will begin in order to figure out the {\color{black}actual} handover time. In the ready-made LTE handover simulations on SimuLTE, the UE sends the handover begining warning to the \textit{s-BS} and the \textit{s-BS} sends the UE information to the \textit{t-BS}. The \textit{t-BS} responds with acknowledgment. After the \textit{t-BS} informs the core network about path switching, the handover process is completed. In our proposal, {\color{black}the} UE starts the handover process with the \textit{t-BS}. The \textit{t-BS} informs the core network and \textit{s-BS} after authentication control. The distance between the BSs is the same for both environments, and the transmission power for BSs and UEs are left at the SimuLTE default values. The time for the UE to initiate the handover operation with the \textit{t-BS} and the \textit{t-BS} to inform the core network is the same for both simulations. Data transfer time between BSs is the main difference between the two simulations.
}

According to the simulated scenario, a terrestrial BS provides service to UEs inside a high capacity football stadium. Due to the {\color{black}excessive} number of UEs, the BS cannot provide the desired QoS. More than one UxNB is sent to the zone throughout the game for capacity injection. According to the scenario, it is necessary to authenticate UxNBs and to handover UEs from terrestrial BS to the nearest UxNB.

\subsection{Capacity Injection}

In parallel with the technological advancements in mobile networks, the peak data rates of downlink and uplink increase. While the average downlink value provided today in LTE technology is $100$ megabits per second (Mbps), the uplink value has been $50$ Mbps \cite{lteuplink}. A BS that encounters a request above this uplink and downlink threshold values will start dropping packets. As a result, there will be a decrease in the QoS values, {\color{black}which} are determined by the service provider.

\begin{figure}[h!]
\centering
\includegraphics[width=\linewidth]{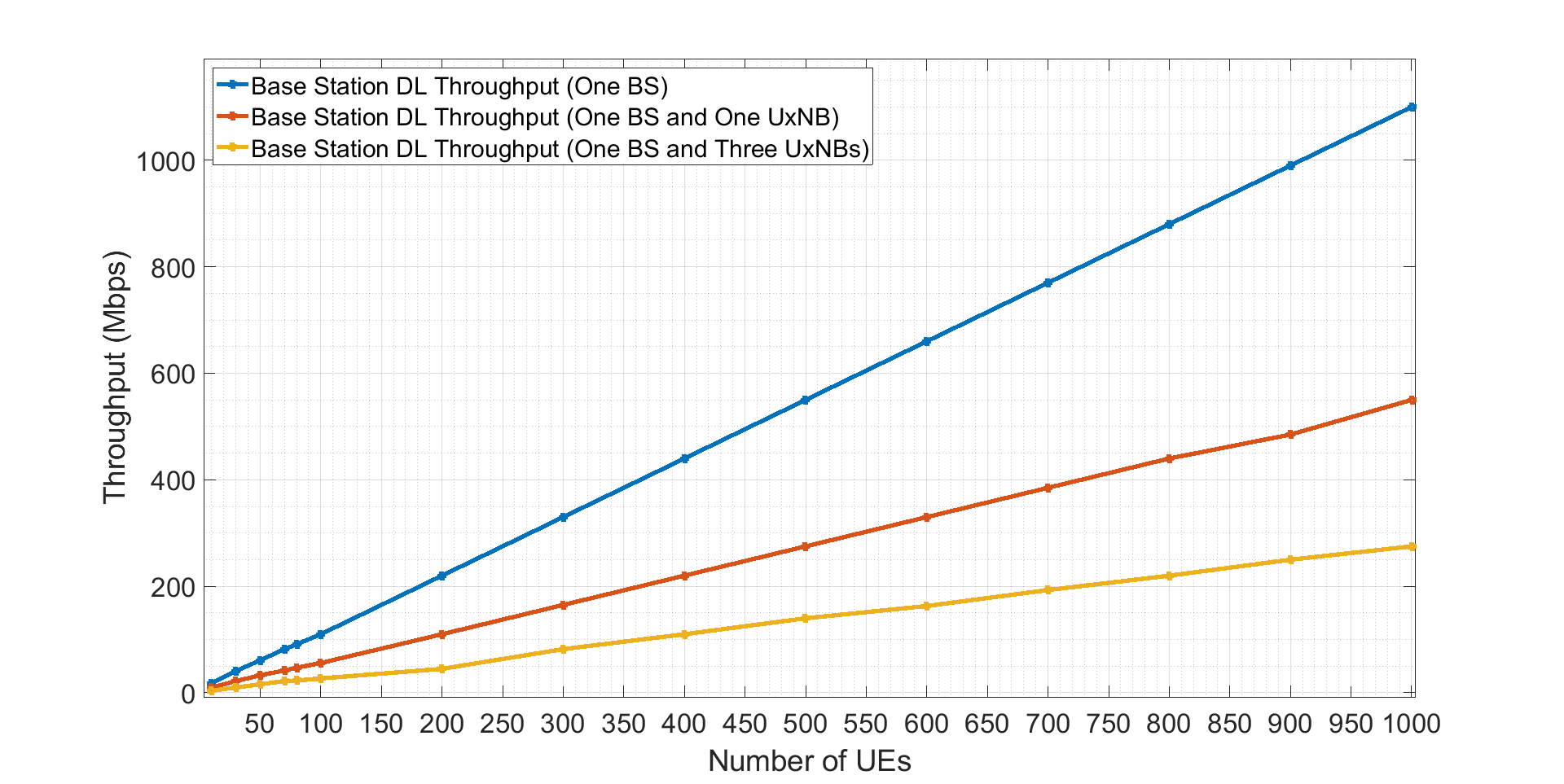}
 \caption{Base station throughput per UE. The throughput is proportional with the number of UEs. The graphic shows that the capacity injection is crucial when the number of UEs is high. {\color{black} The throughput can be reduced with capacity injection by using UxNB.}}
\label{fig:CapacityInjection}
\end{figure}

{\color{black}It is expected that the number of IoT devices connected to the network will reach $60$ billion in $2022$\cite{IoTNumber}. The increase in the number of end devices will also cause increase in the number of groups. Therefore, high-density areas will be encountered more frequently. In high-density areas, such as stadiums, the uplink value will typically be high. In the simulations, it is simulated that UEs request service to watch a video simultaneously. According to the simulations implemented by} SimuLTE, as the number of UEs increases, the required uplink value {\color{black}also} increases, as shown in Figure \ref{fig:CapacityInjection}. For example, the request created by $100$ UEs at the same time creates an uplink value of $110$ Mbps for the BS. Only $100$ UEs can consume all the downlink limit for one terrestrial BS if they all watch video simultaneously.

As can be seen, it is not possible to provide service with only one BS in crowded environments. The use of UxNB emerges as a promising solution. Another question at this point is how many UxNBs on average can be sufficient to cover a stadium. According to the study \cite{droneuplink}, the downlink value for UxNB is $160$ Mbps, a typical flight height of $150$ m. The solution will need one UxNB for approximately $10$ UE. Using too much UxNBs will cause several handover processes. Therefore, a group handover is a promising solution for our scenario.

\subsection{Group Handover}
The latency is one of the main issues in the handover schemes. If the latency is {\color{black}high} in the communications, the quality of service dedicated by providers will be low. The time used up in the handover process, and the number of control packet transmissions between UEs and BS bring along the latency for handover. {\color{black}Recall that in our scenario} several UEs in a football stadium will switch their access network from a terrestrial BS to an UxNB. The pre-designed scheme in SimuLTE, in accordance with standards are exploited to simulate the LTE handover scheme. We reconfigured some of the codes in the LTE scheme in order to attain statistical information about total handover time and the number of control packet transmissions created by BS and UEs. 

\begin{figure}[h!]
\centering
\includegraphics[width=\linewidth]{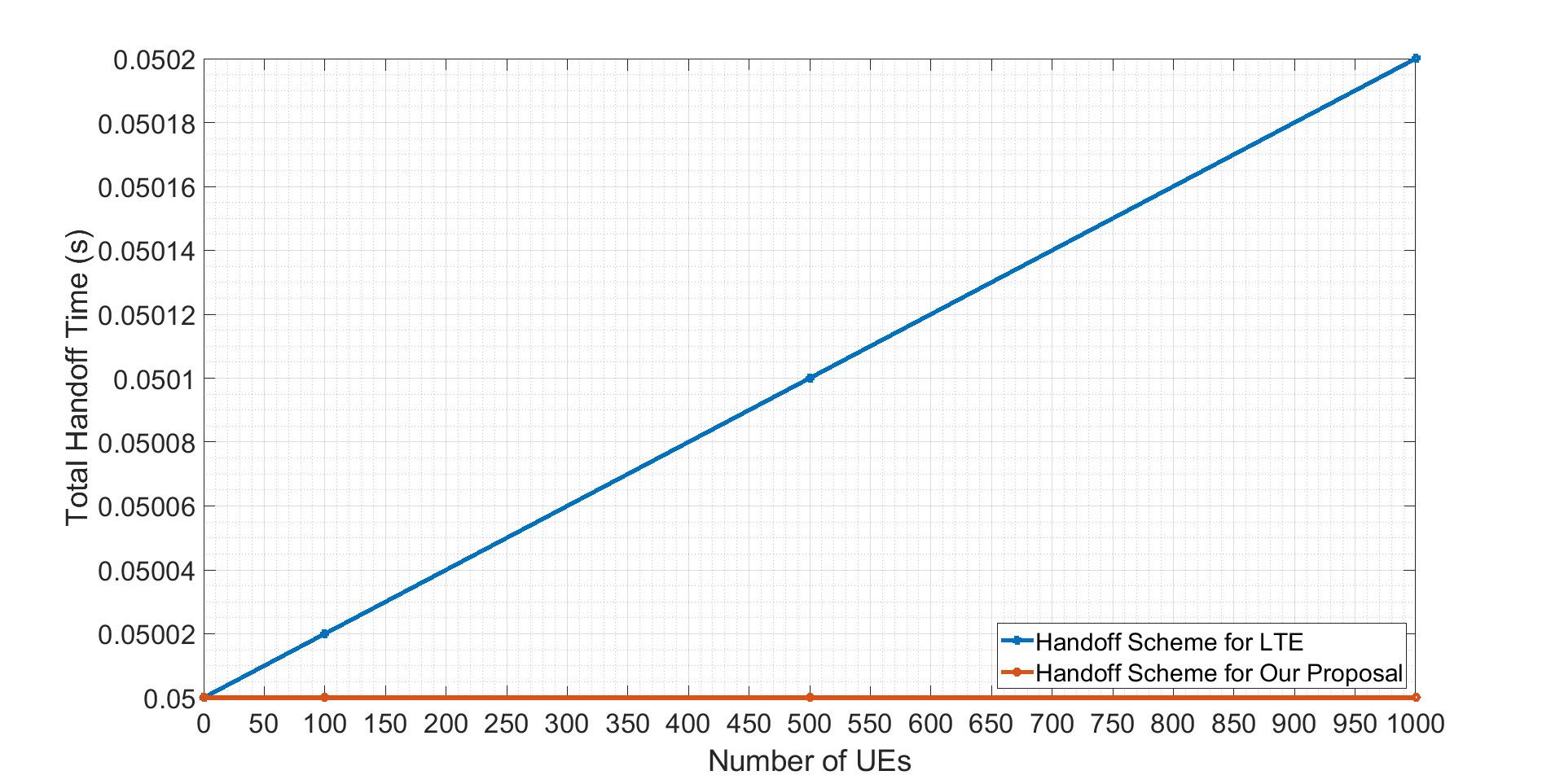}
 \caption{The comparison of handover time. In the proposed method, the handover time is not related with the number of UEs due to the group handover solution. However, in LTE, handover time is proportional with the number of UEs.}
\label{fig:handofftime}
\end{figure}

\begin{figure}[h!]
\centering
\includegraphics[width=\linewidth]{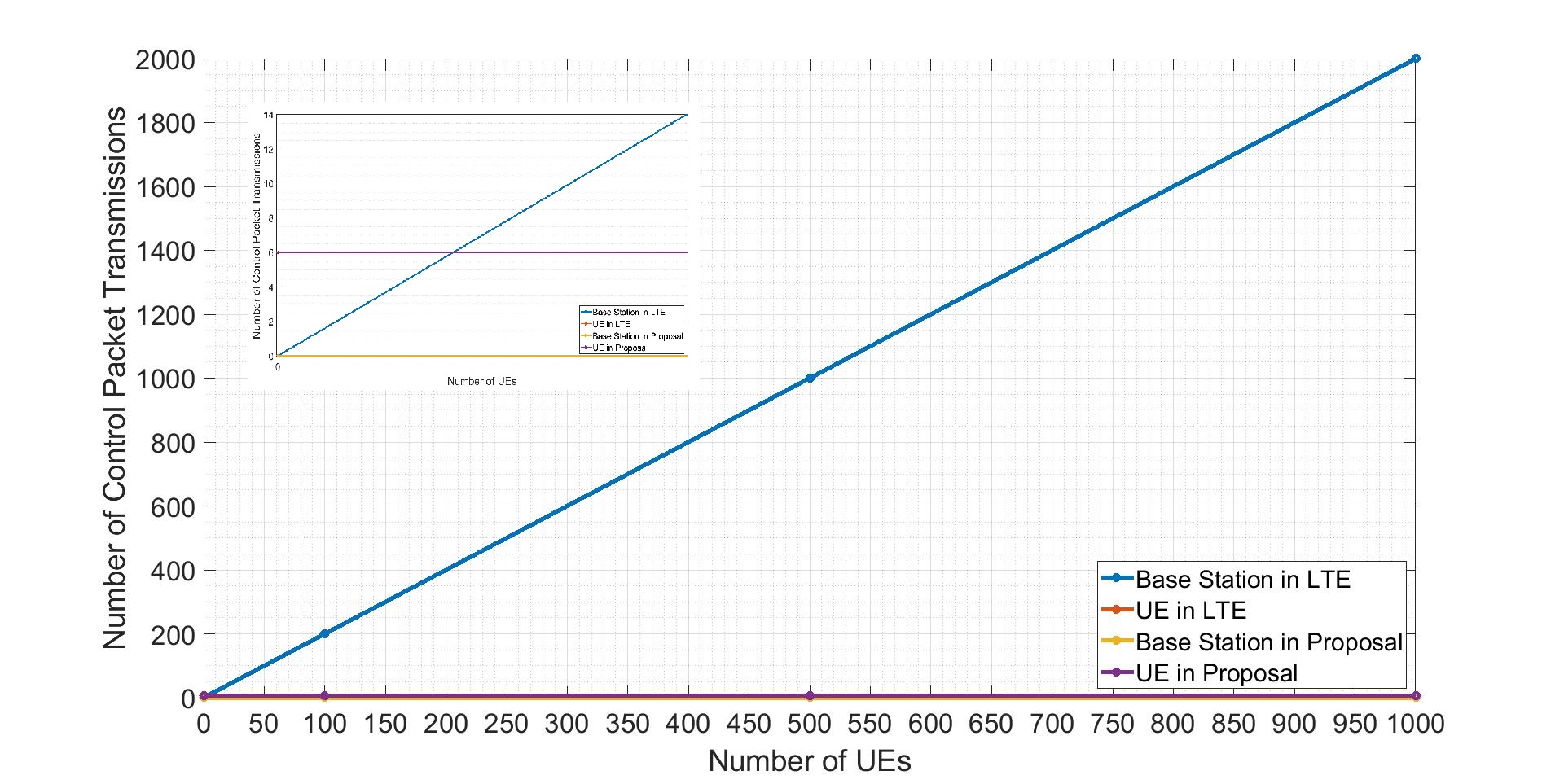}
 \caption{Total number of control packet transmissions. The number of control packet transmissions both in LTE and proposed method is six for each UE. UE needs six control packet transmissions to update the core network. There are no control packet transmissions between BSs for handover in our proposed method. However, in LTE, the \textit{s-BS} must send credentials for each UE to the \textit{t-BS}.}
\label{fig:Transmission}
\end{figure}

As {\color{black}given} in Figure \ref{fig:handofftime}, the total handover time is increasing in the LTE scheme when the number of UEs {\color{black}surges}. The \textit{s-BS} should send user-related data for each UE to the \textit{\color{black}t-BS} according to the standards. Hence, the communication between BSs is proportional to the number of UEs, as in Figure \ref{fig:Transmission}. Each UE is linked with the core network to update handover parameters, and six transmissions from UE to the core network is a fixed value in both standards and our proposal. The most energy-consuming transmissions occurs between BSs. 

{\color{black}As seen in Figure \ref{fig:handofftime},} the number of UEs is not affecting the total handover time. Because a group handover scheme is performed by the \textit{t-BS}. The \textit{t-BS} collects public values of UEs and compares the received values with values produced by the private function. The number of control packet transmissions for our proposal is in Figure \ref{fig:Transmission}. The control packet transmissions per UE is still six as in standards. Because the UE must update handover parameters with the core network. The advantage of our proposal on the LTE is the communication between BSs {\color{black}being} zero. The \textit{t-BS} can handle authentication {\color{black}of} UEs by confirming their public keys without the requirement of communication with the \textit{s-BS}.

\subsection{Performance Assessment}

Both in TS 36.300 \cite{3GPP36300} and our proposal, the number of control packet transmissions {\color{black}for path switching} per UE is six. The reason behind that is the UE contacts with the core network six times to transfer the new connection parameters for the \textit{t-BS}. 

In {\color{black}release-16} \cite{3GPP36300}, the \textit{s-BS} sends the connection information for related UE to the \textit{t-BS}. {\color{black}Each} connection between the \textit{s-BS} and the \textit{t-BS} has an acknowledgment message to endorse the receiving of the data. At the end of the handover process, the \textit{t-BS} informs the \textit{s-BS} for completing the handover. Hence, the total number of control packet transmissions by the \textit{s-BS} or the \textit{t-BS} is {\color{black}twice} the number of UEs. 

In our proposal, the communication between the \textit{s-BS} and the \textit{t-BS} is not performed. The \textit{t-BS} {\color{black}gets} the secret $f(x)$ function, which the key factor for the confirmation, when the \textit{t-BS} becomes active. The \textit{s-BS} (terrestrial) authenticates the \textit{t-BS} (UxNB) when the \textit{t-BS} become active at the football stadium. Once authentication is confirmed, the \textit{s-BS} shares the secret function with the \textit{t-BS}. In the handover process, the \textit{t-BS} performs the confirmation by using the private function. The relevant UE sends the public keys $(x_i,f(x_i)P)$ to the \textit{t-BS}. The \textit{t-BS} performs addition for each $x_i$ value and elliptic curve addition for each $f(x_i)P$ value. Once all the UEs in the group send public values, the \textit{t-BS} compares the total $x_i$ and total $f(x_i)P$.

{\color{black}The total handover time for LTE and our proposal is compared}, the {\color{black}surge} of the number of UEs does not change the total handover time in our proposal. However, the handover time is proportional to the number of UEs in standards. The reason for that is the communication between the \textit{s-BS} and the \textit{t-BS} getting increased if the number of UEs is too much.

{\color{black}
According to the simulation results, the time for one control packet transmission between BSs is approximately $100$ nanoseconds. The \textit{s-BS} sends one packet to the \textit{t-BS} for {\color{black}indicating} information and receives one packet from the \textit{t-BS} for {\color{black}the} acknowledgment. The total time to send one UE data from the \textit{s-BS} to the \textit{t-BS} is $200$ nanoseconds. $0.05$ seconds is the standard time for both our proposal and LTE standard.  This time slot is required to start the handover process by UE and to update the core network about cell change. The reason for the change in handover time in LTE is data sharing between BSs. {\color{black}The} data sharing process for one UE is $200$ nanoseconds, it is $0.2$ milliseconds for $1000$ UEs, and this value increases {\color{black}linearly} as the number of UEs increases, as {\color{black}shown} in Figure \ref{fig:handofftime}.
}
\section{Conclusion}
A group handover solution is proposed in this paper. The main objectives of our study are to decrease the total handover time in the high-density places, to propose an authentication scheme between BSs and to decrease energy consumption by restricting the number of control packet transmissions in the handover process. 

{\color{black}Simulation results} show that the handover time is constant in our proposal, which is {\color{black}$0.05$ seconds}, and there are no control packet transmissions between BS. Whereas according to 3GPP Release 16 \cite{3GPP36300}, the \textit{s-BS} must send data to the \textit{t-BS} for each UE to complete the handover successfully. The handover time and the number of control packet transmissions are entirely related to the number of UEs in the handover process. Hence, the handover time and the number of control packet transmissions between BSs are proportional to the number of UEs.

In our proposal, the \textit{t-BS} can authenticate a group of UEs with  a private function without revealing any security parameter. The discrete logarithm problem in the elliptic curve groups hides the private keys of UEs. The control packet transmission between the \textit{s-BS} and the \textit{t-BS} is zero and not affecting the handover time. Besides, the performance issue of the $X_n$ link between terrestrial and UxNB can be solved by our authentication solution.

The scenario for which we propose our scheme does not include a high mobility. The terrestrial and UxNB are both fixed. The UxNB comes over the high-density area and provides service to the UEs in its coverage area. The main objective of UxNB is to decrease the burden on the terrestrial BS. However, UxNB is going to move into urban or disaster areas. 

In case of a high number of UxNBs in the air to cover all the areas along with densely deployed UEs, new schemes are required to provide authentication between several UEs and UxNB, between UxNBs. {\color{black}Additionally}, UEs will be transferred from one to another UxNB. Handover schemes are also {\color{black}needed for the beyond-5G networks in the future.}

{\color{black}
5G is currently dominating mobile networks around the {\color{black}globe} and researchers have  already started 6G studies. Regarding security, it is predicted that 6G will face with more challenging problems \cite{6G}. Artificial intelligence should be used actively to {\color{black}address} security problems in mobile networks. Since end-to-end security will be more important for 6G due to the increasing of decentralized UEs, {\color{black}implementation} of security protocols with artificial intelligence {\color{black}has the potential to offer significant benefits with the introduced autonomy}. Due to the increase in peer to peer communications, authentication solutions require decentralized proposals instead of central solutions. The use of blockchain technology should {\color{black}also} be included in authentication schemes because of {\color{black}its straightforward} deployment for distributed users. There is no need for a trusted party in blockchain-{\color{black}based approaches}.
}

\begin{IEEEbiography}[{\includegraphics[width=1in,height=1.25in,clip,keepaspectratio]{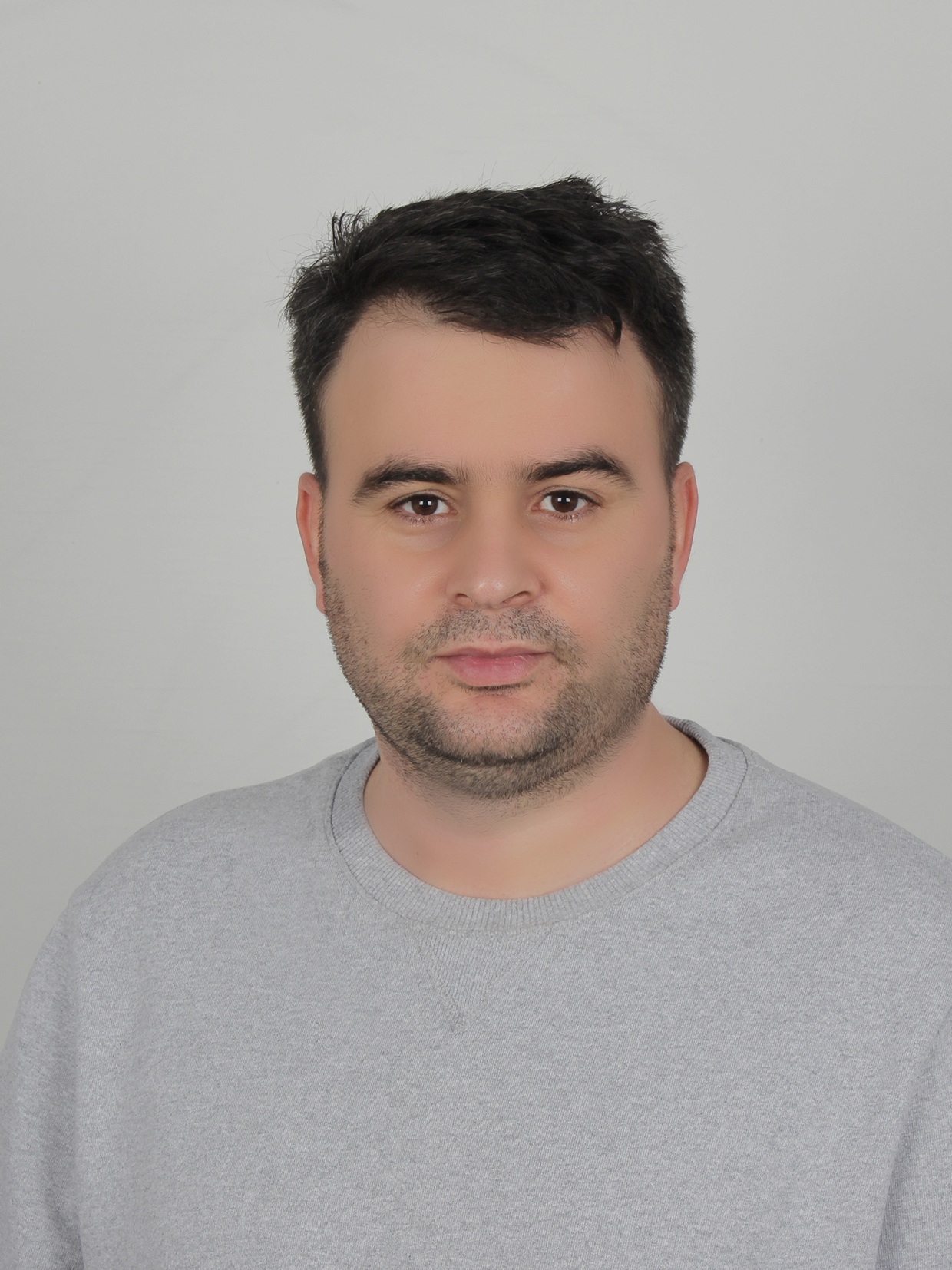}}]{Yucel Aydin}
(aydinyuc@itu.edu.tr) received the M.Sc. degree in cyber security engineering and cryptography program from Informatics Institute, Istanbul Technical University, Istanbul, Turkey, in 2017, where he is currently pursuing the Ph.D. degree.

His research interests are network security, cryptography and computer security.
\end{IEEEbiography}

\begin{IEEEbiography}[{\includegraphics[width=1in,height=1.25in,clip,keepaspectratio]{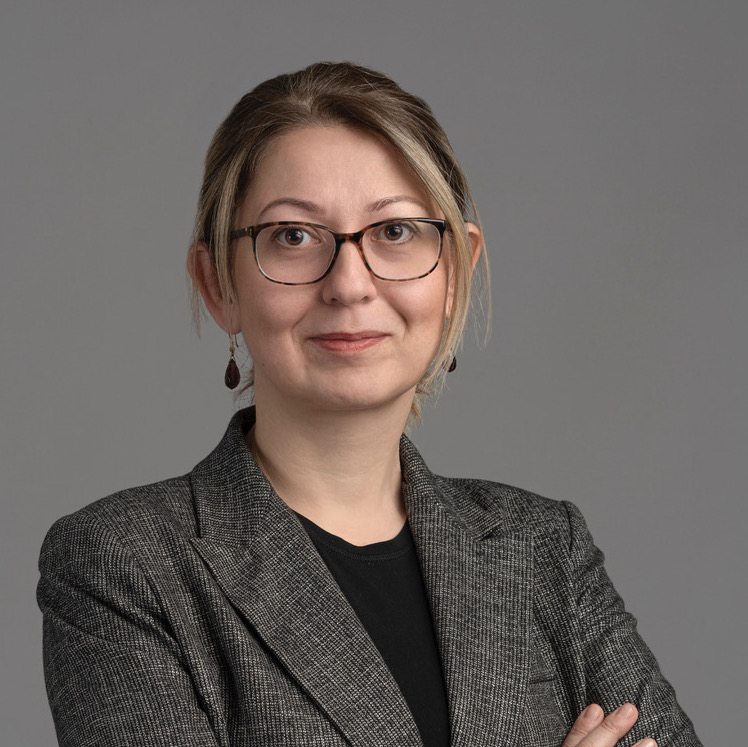}}]{Gunes Karabulut Kurt}
[StM’00, M’06, SM’15] (gunes.kurt@polymtl.ca) is currently an Associate Professor of Electrical Engineering at Polytechnique Montréal, Montreal, QC, Canada.  She received the B.S. degree with high honors in electronics and electrical engineering from the Bogazici University, Istanbul, Turkey, in 2000 and the M.A.Sc. and the Ph.D. degrees in electrical engineering from the University of Ottawa, ON, Canada, in 2002 and 2006, respectively. From 2000 to 2005, she was a Research Assistant at the University of Ottawa. 

Between 2005 and 2006, Gunes was with TenXc Wireless, Canada. From 2006 to 2008, she was with Edgewater Computer Systems Inc., Canada. From 2008 to 2010, she was with Turkcell Research and Development Applied Research and Technology, Istanbul. Gunes was with Istanbul Technical University from 2010 to 2021. She is a Marie Curie Fellow and has received the Turkish Academy of Sciences Outstanding Young Scientist (TÜBA-GEBIP) Award in 2019. She is an Adjunct Research Professor at Carleton University. She is also currently serving as an Associate Technical Editor (ATE) of the \textit{IEEE Communications Magazine} and a member of the IEEE WCNC Steering Board.
\end{IEEEbiography}

\begin{IEEEbiography}[{\includegraphics[width=1in,height=1.25in,clip,keepaspectratio]{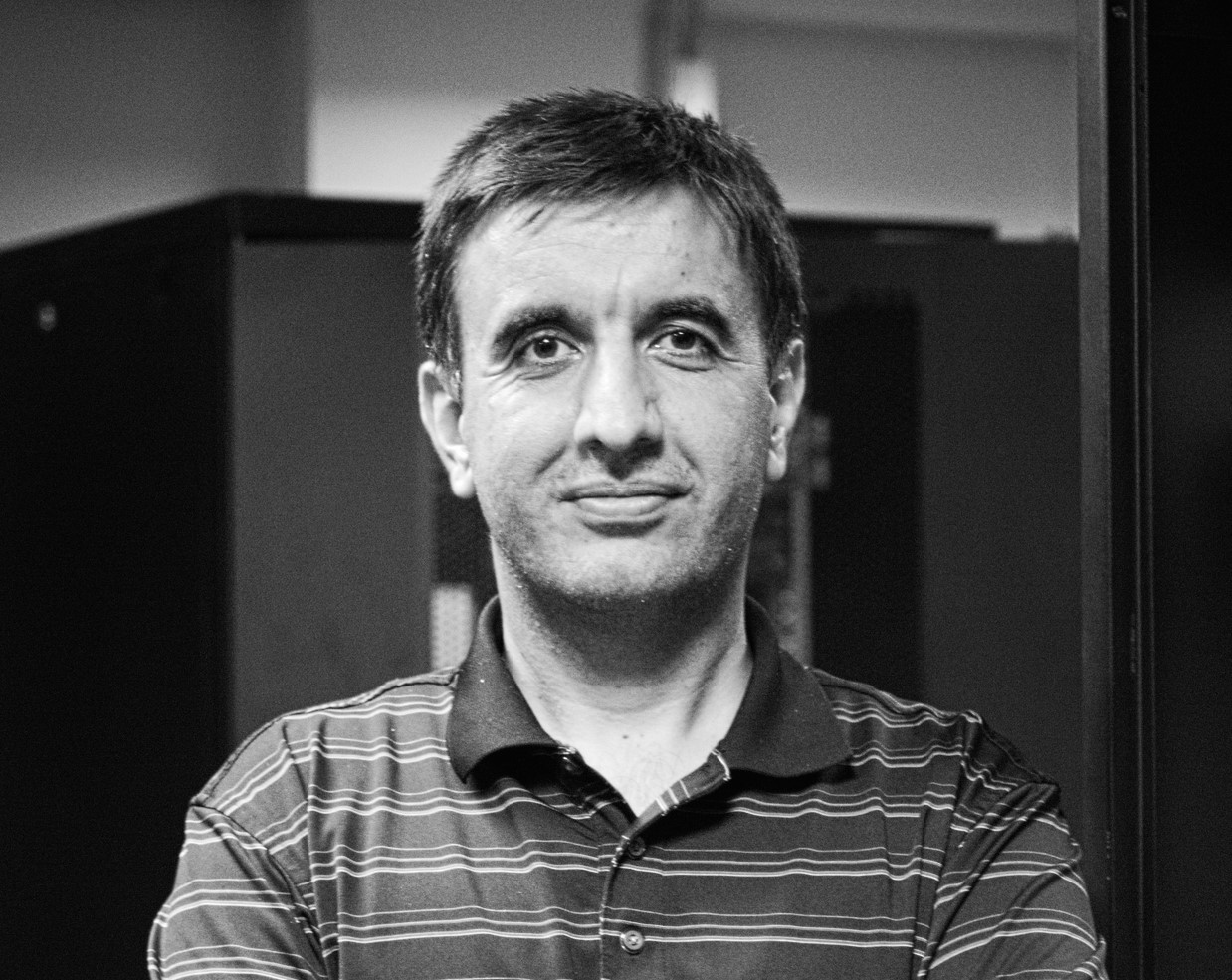}}]{Enver Ozdemir}
(ozdemiren@itu.edu.tr) received a Ph.D. degree in mathematics from the University of Maryland, College Park, MD, USA, in 2009. 

He is currently an Associate Professor at Informatics Institute, Istanbul Technical University, Istanbul, Turkey. He was a member of the Coding Theory and Cryptography Research Group, Nanyang Technological University, Singapore from 2010 to 2014. His research interests include cryptography, computational number theory and network security.
\end{IEEEbiography}

\begin{IEEEbiography}[{\includegraphics[width=1in,height=1.25in,clip,keepaspectratio]{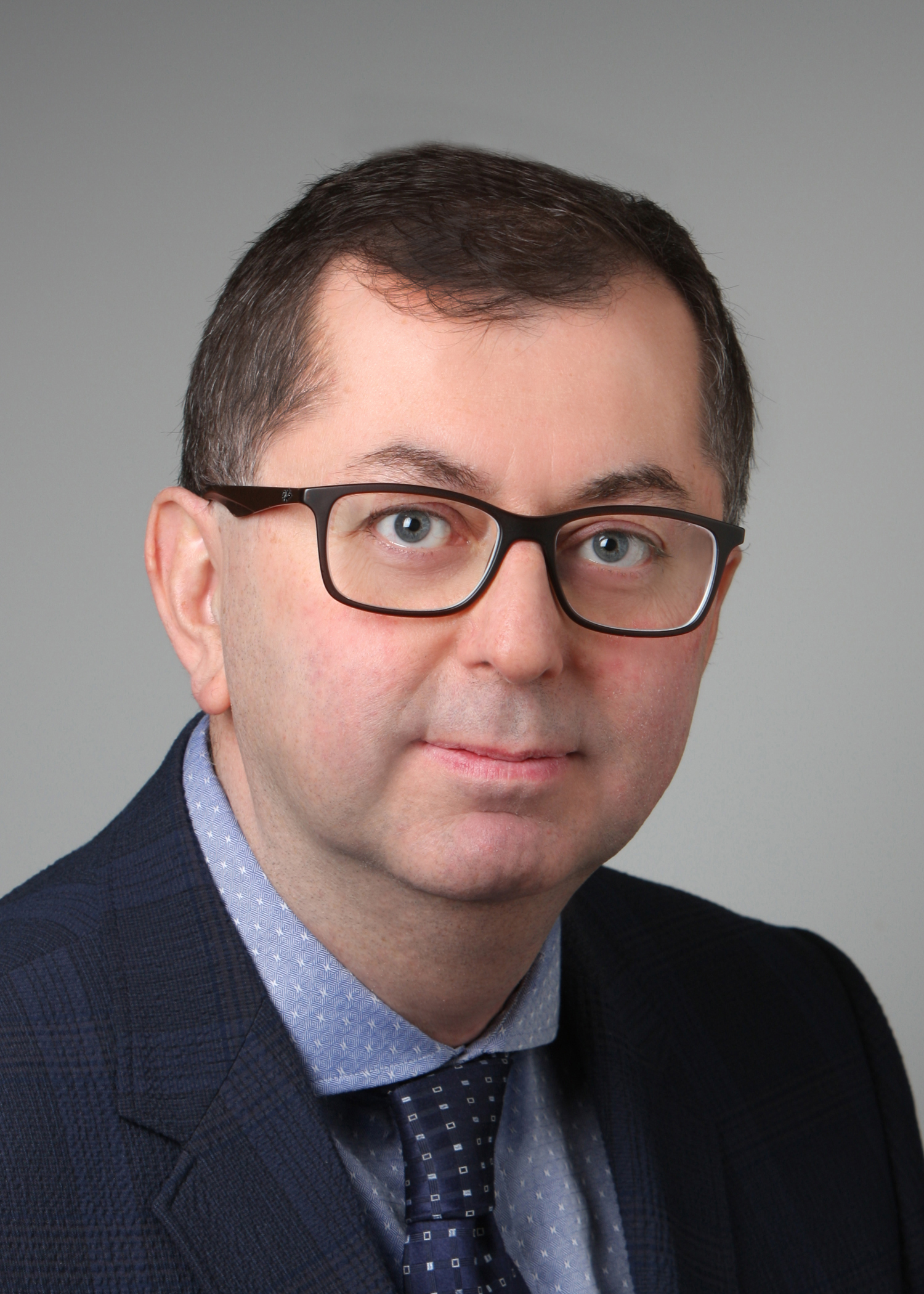}}]{Halim Yanikomeroglu}
[F] (halim@sce.carleton.ca) is a Professor in the Department of Systems and Computer Engineering at Carleton University, Ottawa, Canada. His primary research domain is wireless communications and networks. His research group has made substantial contributions to 4G and 5G wireless technologies. During 2012-2016, he led one of the largest academic-industrial collaborative research programs on pre-standards 5G wireless. In Summer 2019, he started a new large-scale project on the 6G non-terrestrial networks. His extensive collaboration with industry resulted in 37 granted patents. 

He has formally supervised or hosted at Carleton a total of 135 postgraduate researchers in all levels (PhD \& MASc students, PDFs, and professors). He has coauthored IEEE papers with faculty members in 80+ universities in 25 countries and industry researchers in 10 countries. He is a Fellow of IEEE, EIC (Engineering Institute of Canada), and CAE (Canadian Academy of Engineering), and a Distinguished Speaker for both IEEE Communications Society and IEEE Vehicular Technology Society. 

He is currently serving as the Chair of the IEEE WCNC (Wireless Communications and Networking Conference) Steering Committee. He was the Technical Program Chair/Co-Chair of WCNC 2004 (Atlanta), WCNC 2008 (Las Vegas), and WCNC 2014 (Istanbul). He was the General Chair of IEEE VTC 2010-Fall (Ottawa) and VTC 2017-Fall (Toronto). He also served as the Chair of the IEEE’s Technical Committee on Personal Communications. Dr. Yanikomeroglu received several awards for his research, teaching, and service, including the IEEE Communications Society Wireless Communications Technical Committee Recognition Award in 2018 and IEEE Vehicular Technology Society Stuart Meyer Memorial Award in 2020.
\end{IEEEbiography}

\end{document}